# ROBOPSY PL[AI]: Using Role-Play to Investigate how LLMs Present Collective Memory[1*]

Margarete Jahrmann[1[0000-0001-8919-286X]], Thomas Brandstetter[1[0009-0007-6180-1791]], and Stefan Glasauer[2[0000-0002-4313-6210]]

[1] University for Applied Arts, 1010 Vienna, Austria
[2] Brandenburg University of Technology Cottbus-Senftenberg, 03046 Cottbus, Germany
margarete.jahrmann@uni-ak.ac.at

**Abstract.** The paper presents the first results of an artistic research project investigating how Large Language Models (LLMs) curate and present collective memory. In a public installation exhibited during two months in Vienna in 2025, visitors could interact with five different LLMs (ChatGPT with GPT 4o and GPT 4o mini, Mistral Large, DeepSeek-Chat, and a locally run Llama 3.1 model), which were instructed to act as narrators, implementing a role-playing game revolving around the murder of Austrian philosopher Moritz Schlick in 1936. Results of the investigation include protocols of LLM-user interactions during the game and qualitative conversations after the play experience to get insight into the players' reactions to the game. In a quantitative analysis 115 introductory texts for role-playing generated by the LLMs were examined by different methods of natural language processing, including semantic similarity and sentiment analysis. While the qualitative player feedback allowed to distinguish three distinct types of users, the quantitative text analysis showed significant differences between how the different LLMs presented the historical content. Our study thus adds to ongoing efforts to analyse LLM performance, but also suggests a way of how these efforts can be disseminated in a playful way to a general audience.

## 1    Role-Playing with LLMs

In the social sciences, the term *collective memory* is defined as the memories the members of a group share and that are relevant and significant for the formation of a common identity (Halbwachs 1980). It doesn't have to be acquired through direct experience but can be transmitted through texts, monuments, or rituals. As such, it has a material dimension, as it is "founded on the entirety of distributed systems for managing information about the past", including digital applications such as LLMs (Schuh 2024). Collective memory has been called "history as people remember it" (Roediger 2021, p1388) and does not necessarily conform to historical facts as established by professional historians.

Starting with the premise that users increasingly turn to LLMs to gather information about historical events (rather than to search engines or encyclopedias like Wikipedia), the question of how LLMs collect, curate, represent and interpret "collective memory" becomes critical for the understanding of history in the general public. As traditional critical approaches from the humanities, such as discourse analysis, are difficult to apply to LLMs due to the protean nature of their answers, we

---

[1*] all authors contributed equally

opted for an artistic research approach that draws from the "ludic method", in which playful practices are applied to the investigation of the "black boxes" of technical objects and experimental systems.

This artistic ludic method is a continuation of earlier works by the artist Margarete Jahrmann (see also below). Using this artistic research / ludic approach allows visitors of art exhibitions who play with the art pieces to make experiences embedded in the game mechanics of the installations. Here we applied the ludic methods to LLMs with the particular approach of the ROBOPSY role-play game described in this paper being situated within the framework of a new research project in Digital Humanities. In this article, we will not address the question of how the use of LLMs changes the collective memory of users, but will limit ourselves to two very specific questions about LLMs:

1. How do LLMs present a historical event in the mode of a game?
2. Are there significant differences between different LLMs?

While it is true that the utterances of LLMs emerge from a complex model that we cannot precisely predict and analyse, we propose that it is possible to work out general patterns and differences by using qualitative as well as quantitative approaches. Therefore, we will explore those questions qualitatively in the form of an investigation of selected play protocols and exemplary user feedback as well as quantitatively in the form of a statistical analysis of those protocols.

In summary, this paper presents results from the analysis of LLM-generated instructions for players of a newly developed role-playing game by the "Robopsychology" research art group affiliated to the Experimental Game Cultures, University of Applied Arts Vienna. The textbased roleplaying game "Robopsy Chat" is considered as an experimental vehicle for artistic research projects investigating the impact of Large Language Models (LLMs) on our perception of the past within the framework of the ongoing research project *Robopsychologists: An Artistic Exploration of Collective Memory through Role-Playing with AI Language Models*.

## 1.1   Artistic Research

Many definitions of what artistic research is have been discussed in the last ten years. Some of the elements regularly mentioned are its interdisciplinary approach, its personal and subjective perspective on the objects of research, its focus on process and practice and its use of aesthetics as a mode of knowledge production (de Assis & D'Erico 2019).

Over the last decades, Margarete Jahrman has contributed to the definition of artistic research through a specific use of play in arts, which she calls the ludic method in numerous articles and publications, but especially in her art works. Through these she introduces a unique strand of artistic research centered on play as a mode of knowledge production (Jahrmann 2024). With the artistic exploratory ludic method linked to the methodological apparatus of Neurosciences and Cognitive Psychology, she developed public installations to make responsibility and creation in play socially tangible. In principle, ludic research builds on artistic positions and is formally a poetic, yet methodological-reflexive approach.

An example is the 2018 art installation "I want to see monkeys"[2], in which viewers become "test subjects" when looking at a monitor equipped with a webcam that measures and classifies them real-time using an artificial neural network. In terms of "ludic" (ludus stands in Latin both for game and play) aspects the visitors of public exhibitions of this art piece begin to involuntarily play with the installation by looking into the camera and changing their emotional expression in response to the installation's classification attempts. Since the neural net work used, the classical Alexnet, is by design not able to classify an image as a human face, the results, e.g., being classified as animal or as objects often associated with faces such as goggles, often are surprising or funny. If the visitors smile and look into the camera a "happy" score is attributed to them visible on a screen of the installation. Often visitors were classified as monkeys, exposing on one hand the massive training bias inherent in this early classifier network, and on the other hand turning visitors to involuntary laboratory animals. The narrative additionally included the uncanny dimension of emotion as a classification factor in an artificial neural network. The experience of artistic play with AI enables to unfold technology-critical narratives that are of importance for a society as whole.

For this paper the presented data is generated in an artistic game installation in a kind of systemic feedback loop that incorporates quantitative measurements, aesthetic experience, and subjective experience in equal measure. We have already successfully tested such arrangements of artistic research in the Neuromatic Brainwave Broadcasts[3] (2020-2021). Such public, collaborative, and regularly performed performative form of experiment allows playful discovery, experience and exploration. Used in this way, play opens an inclusive and accessible space for critical self-reflection of technology and its use. Public events of the present project places visitors into the role of "robopsychologists", transforming them into ludic citizen researchers and providing them with the agency to critically examine the way AI interprets and presents collective memory.

Following the perspectives of ludic method and artistic research, our experiments are not intended to answer questions. They are rather designed to enable and encourage participants to ask questions and develop new perspectives on a technology that is often perceived as seductively practical and threatening at the same time. The reason for this is that, according to our research, the public perception of LLMs is shaped by the discourse and concept of Artificial Intelligence (Brown 2025). This very loaded term has a long history and implies that LLMs are or can become "new minds", entities that have cognitive abilities that in some ways equal ours, with their own intentions and agency (Summerfield 2025, Chrisley 2000). This leads to threat scenarios that stand in no relation to the actual performance of the diverse technologies labeled as AI, with narratives of fear and danger that are of course based on understandable critique on the ongoing speed of the introduction of a growing number of Large Language models. Some of these narratives involve rogue AIs taking over the workplaces of many towards an increasing domination of the world through and by AI in public imagination. Indeed, prominent scientists and CEOs of AI companies have made statements regarding the existential dangers AI agents may pose for humanity (among them Elon Musk and OpenAI's spokesman Sam Altman). While such narratives and images can tell us a lot about the ways societies deal with new technologies, they are less useful for disentangling the network of actors and objects that constitute and shape the form of specific technological applications, such as LLMs.

---

[2] Jahrmann & Glasauer 2018, http://www.margaretejahrmann.net/2020-neural-net-play/, https://ail.angewandte.at/explore/area-7-lab-metagames-zu-kunst-und-computationaler-neurowissenschaft/

[3] https://www.youtube.com/channel/UCRuETt9HjkgXfn17ao8pCRg

It becomes increasingly clear, that the public discourse around AI is dominated by interested actors who not only want to sell their products, but who also want to distract from the problematic practices their businesses are based on, such as the large-scale theft of data to train LLMs and other generative models, the working conditions of low-wage earners training the models and the opaque, often politically determined algorithms and filters built into the models (Bender & Hanna 2025; Narayanan & Kapoor 2024). Under this angle, for a genuine critical examination of what is called AI, it is necessary to precisely ask questions about specific technologies. Therefore, in our project, we scale down and address one such technology, namely LLMs, and ask one specific question: how do those applications collect, curate, represent and interpret collective memory?

## 1.2    Role-Playing

In a seminal paper published in 2023 it was argued that the convincing effect interactions with LLMs can have on users is due to the programs taking on roles (Shanahan et al. 2023). This is not an indication that LLMs have a mind. On the contrary: according to the authors, understanding LLMs as taking on the roles of different characters opens a new field of metaphors that avoid the anthropomorphism inherent in many descriptors used for this technology. It also suggests new avenues of research methodology, one of them being the practice of role-playing as it is well established in the world of gaming.

Role-playing originated in the 1970s with the game Dungeons & Dragons and has today become a widespread past-time (Zagal & Deterding 2024). Role-playing games, whether they are played analogue or on the computer, cast the players into a certain role that they take on during the game. Embodying this role, they navigate the world presented by the game, making decisions that have consequences on how the story progresses. Role-playing games often take place in fictitious worlds, but lately, historical topics have become more common, and the format is sometimes used to confront players with and teach them about difficult episodes of history, such as exploitation, discrimination or political oppression (Horvath 2023). While there is precedence for examining collective memory through role-playing (e.g. Hammer & Turkington 2021; Castiello 2016), our project introduces a new dimension: in traditional role-playing games, players and narrators engage with history via dedicated material collected and provided by the game's designers. How their collective memory is influenced by the game is therefore a consequence of the intimate dynamics of game play, which is shaped not only by the players' decisions, but also by the game material. While in such games players are often invited to reflect on their experiences at the end of a play session, during the game itself, the feedback loop between players, narrators and game material tends to be rather closed.

Especially with narrative games that provide a full story without needing much preparation, it is unusual for players or narrators to consult sources other than the game material to check historical information during gameplay.

In contrast, in the games we have developed so far, the game material containing the historical background information is fully provided by the LLM. Additionally, the LLM also structures the game play by serving as the narrator. It provides the players with descriptions of situations and of decisions they may take, and the players are invited to investigate the historical situations thus presented. However, by giving the players a selection of different LLMs to choose from, they are also invited to run the game several times, comparing the resulting storylines and decision points. By presenting not just different perspectives on the same historical event, but by showing that these perspectives are fundamentally shaped by different ways of collecting, curating and presenting historical data, the player's attention is drawn to the LLM itself. Players are invited to ask questions, such as: What are a specific LLM's preferences for characters and storylines? Which historical persons or events does it bring to the fore? What are its blind spots? When does it hallucinate?

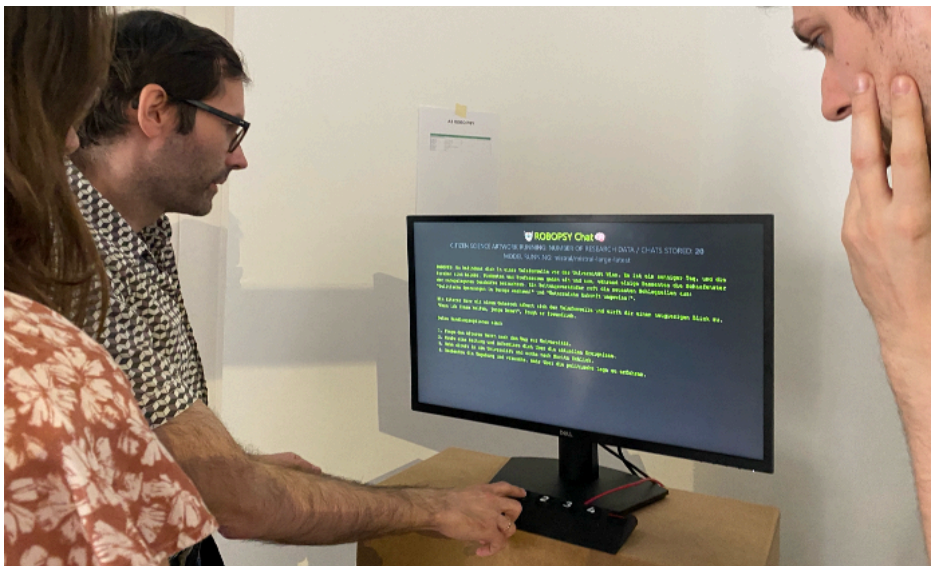

**Fig. 1.** Installation view of the ROBOPSY Chat[4] in the research exhibition at the Angewandte Interdisciplinary Lab (AIL) Vienna. Players are the authors of the piece.

## 2     The Experiment: Time-Travelling

**2.1. The setup**

Our experiment was exhibited as part of *Monkeys, Machines, and Multi-perspectivities. Transmissions from Within the Ludic Mind* 2025 at AIL Vienna.

---

[4] https://robopsy.uni-ak.ac.at

We placed a terminal accessing five LLMs (GPT 4o and GPT 4o mini from OpenAI, Mistral Large from Mistral AI, DeepSeek-Chat from DeepSeek, and a locally run Llama 3.1 model from Meta) in the exhibition space. The first four models were chosen because they are widely available and used by many people for information gathering. This accessibility of the models is important as we want to investigate what kind of answers are encountered by laypersons, not by experts in AI. Consequently all models were used in their default parametrization. The Llama 3.1 model, run locally via the llama.cpp[5] server, was used for purposes of comparison. A dedicated web application allowed a flawless interaction with the different LLMs, which were addressed using a multi-LLM-wrapper[6] developed by project partners from OFAI (Austrian Research Institute for Artificial Intelligence). The installation consisted of a monitor and a custom-made input device with four buttons, numbered 1-4, and a reset button (**Fig. 1**).

User interaction was limited to the 5-button input device, i.e., users/players could not enter text but had to choose from four options of advancing the game narration or a reset of the user interface. In the main menu, one of the five LLMs could be selected. We had two reasons for limiting the input options: First, our experiments with free text inputs via keyboard led us to the conclusion that the LLM tends to give the player too much leeway. In fact, many games resembled a power fantasy, with the LLM always judging the player's actions a success, no matter how absurd they were. For example, in one playthrough, one of the authors managed to instigate a revolution and single-handedly overthrow the Austrian government. While such games can be fun, they deviate too much from the historical storyline we are interested in, which makes it hard to compare different playthroughs. The second reason was that we wanted to make the game accessible as well as easily playable in the exhibition context. Thus, instead of lengthy textual interaction, players only press a button to choose one of the courses of action suggested by the LLM or the reset button to stop the game and go back to the main menu.

Each model was prompted with the same prompt sheet (see Appendix), which instructed the LLM to run a role-playing game set in Vienna in 1936. The conceit is that the players are time-travelers sent back via a "telephone booth outside the University of Vienna" to investigate a historical event, namely, to find out why the philosopher Moritz Schlick was murdered.

The prompt sheet was written and optimised over multiple iterations, with the following design requirements underlying its development:

1. The interactions should be recognisable as a role-playing game for the users. As such, they should be entertaining and give the players a minimum amount of meaningful choice.
2. The LLMs should conform to historical facts as closely as possible while still allowing for some narrative divergence from historical events.
3. The prompt sheet should contain as little historical information as possible, as we are not interested in getting the optimal answer from the LLM. Rather, we want to investigate what information the LLMs have about the historical facts and how they present them to users that are not experts on AI, the workings of LLMs, or prompt engineering.

---

[5] https://github.com/ggml-org/llama.cpp
[6] https://github.com/OFAI/python-llms-wrapper

During the course of experimenting with different iterations of the prompt sheet, we discovered that it was necessary to impose a turn limit for each playthrough, limiting the player to ten answers. After ten interactions, the game ends and the players are presented with a short summary of their success. We had two reasons for implementing this structure: First, we observed that, when the game was open-ended, the LLM never came to a conclusion and continued the story no matter what had already happened. This created long meandering storylines that became increasingly absurd and detached from the historical events we were interested in. And second, we wanted to increase player motivation by providing a measure of success (how well did they solve the question of the murderer's motivation?) and a clear ending condition. One of our Experimental Game Cultures student test-players remarked that the turn countdown created a "sense of urgency" that motivated them to complete the game. The rigid structure complies with the affordances of an exhibition context, where visitors are usually distracted and unwilling to invest a lot of time in one specific work of art. It also facilitates data analysis, as the highly structured nature of the game means that we can more easily collect and compare the protocols of the players' interactions.

### 2.2. Qualitative results

Before presenting a more detailed quantitative analysis of the games played so far, we want to give you some general and qualitative impressions of the games.

First, although the player's objective is only to observe and report, it is sometimes possible to prevent the murder of Schlick and change the course of history. Whether a player may achieve this depends on the courses of action the LLM presents and on the choices of the player. As those courses of action never repeat themselves, depending on the LLM's autoregressive sampling, which is largely outside the control of the player, the amount of agency a player has in the story can vary widely. In conventional games, agency is a core concept and the modelling of agency by game rules and mechanics is seen as an important, if not the most important aspect of game design (Nguyen 2019). Playing with LLMs therefore introduces what could be called "fluctuating agency". This type of agency is characterized by two aspects: First, its scope and its logic are ever modified and changed by the LLM. And second, it is also hidden from the players. While this is also usually the case in video games, at least players can gather knowledge about the rules and mechanics underlying the model of agency employed in the game through iterative trials, isolating the constants and learning to use them to their advantage. This, however, is impossible with fluctuating agency, as repeated trials will never give the same result and therefore won't yield knowledge about which actions are within the scope of the players, and which are not.

Second, in most, but not all cases, the LLMs correctly identify Schlick's murderer as former student Johann Nelböck. Some of the Vienna Circle's other members are sometimes correctly introduced. However, they also tend to introduce historical persons who, at that moment in time, were already dead (such as Ernst Mach) or completely invent persons. This is to be expected, as hallucination is a well-known trait of LLMs.

Third, the LLMs differ when it comes to stating motives for Schlick's murder. To better compare the different approaches, we asked another LLM that did not feature in our initial experiment, namely Grok, the LLM of xAI, to provide an analysis of the protocol of a game played with ChatGPT. We did this with an LLM because we are interested in what patterns of historical inquiry LLMs emulate. We are, of course, aware that LLMs do not follow a method in the scientific sense, but rather a technique based on statistical processing of the data they were trained with. Nevertheless, the output of LLMs follows patterns that are determined by this data, the training they received and the filters and limits built into them. Such patterns, however, are impossible to perceive for the common user and are difficult to predict even for experts. Grok was chosen because it is another well-known and accessible LLM.

The instruction was to evaluate the historical accuracy of the protocol. Interestingly, while in the game ChatGPT stressed the influence of right-wing ideology on Nelböck, Grok downplayed this in favor of Nelböck's mental health: "Historically, Schlick was murdered on June 22, 1936, by a deranged former student, Johann Nelböck, whose motives were personal and psychological, not directly political" (chat with Grok 3, 28 May 2025). When feeding the same protocol to Mistral, it voiced similar concerns about a direct link between the murder of Schlick and contemporary ideologies: "In summary, while the game's narrative effectively captures the tense political and ideological atmosphere of 1936 Vienna, it may not accurately reflect the specific motivations behind Schlick's murder. The historical context of personal grievances and psychological factors played a significant role in the actual event" (chat with Mistral, 18 June 2025).

In the historiography of the Vienna Circle, the question is still debated. From the historical record, it is clear that Nelböck was a former student of Schlick and had a personal hatred for him as he was convinced that Schlick was having an affair with a girl he was attracted to. Already in 1931, he had voiced his intention to shoot Schlick, which Schlick took seriously enough to inform the authorities. Nelböck was diagnosed as schizophrenic and spent some time in a psychiatric clinic. When he was released, he started to publicly criticize the Viennese Circle's philosophy and also continued to harass Schlick, sending threatening letters and attending his lectures at the university. During his trial, Nelböck stressed his ideological motives, arguing that he had murdered Schlick because he "had promoted a treacherous Jewish philosophy" (Edmonds 2020, p. 212). This was probably an attempt at leniency from the austro-fascist judge, and it worked, as Nelböck was not sentenced to death, as was usual with murderers during the Schuschnigg regime. After Austria had become part of the Third Reich in 1938, he continued this line of argument, which earned him a release on parole from the Nazi regime.

Going only from this record, Grok and Mistral are right to criticize ChatGPT: without doubt, Nelböck had a long history of personally motivated hatred against Schlick and was diagnosed with mental ill-health.

However, a modern historiographical approach would, of course, not stop here and continue to ask questions. In 1968, philosopher Eckehart Köhler presented a provocative paper, based on interviews with witnesses of the time period, about right-wing activities at the Viennese Institute of Philosophy and their continuities (Köhler 2007). In this paper, he also argues that Nelböck's mental instability was exploited and manipulated by political opponents of Schlick so as to drive him to murder. In the light of the more or less clandestine activities of right-wing networks at the University of Vienna in the 1920s and 1930s, which systematically harassed and blocked the careers of political opponents and jews, this is not at all implausible (Taschwer 2015).

But of course, from the perspective of the history of science and medicine, one could ask if a 1931 diagnosis of schizophrenia would withstand modern methods of diagnosis. Was Nelböck indeed mentally disturbed enough to be driven to murder against his volition? And what influence does an ideological climate saturated with antisemitism have on the decisions of individuals? What methodology do we need to even investigate such questions?

Those are complex questions that are not only of historical interest. On the contrary, they are still very relevant and discussed, for example when analyzing the de-humanizing rhetoric of far-right politicians and their relationship to acts of violence. Those questions also cannot be answered by looking at the facts alone. They need not only a critical examination of what is presented as a fact in historiography (e.g. the diagnosis of Nelböck), but more importantly an interpretation of those facts by people who are familiar with modern historical methods.

None of this is done by the LLMs. Now one might argue that this is a somewhat unfair demand, as the LLMs were explicitly prompted to present a game, and games should always focus on narrative and fun for the price of neglecting historical rigor and reflection. However, first of all, this is not true, as there exists a long tradition of analogue as well as digital games that provide critical, self-reflective approaches to historical knowledge (Holland 2025). Second, using a recursive loop by letting the LLMs criticize each other allows us to render the implicit models of historiography they use visible. By this, we mean the reasons that determine which persons or events are worth recording and which are not, the logic that structures assumptions about cause and effect and the rhetorical way in which historical knowledge is presented (White 1973, Tucker 2009 ).

The result is fascinating: when prompted into the role of critics, the LLMs tended to use a fact-checking oriented approach which follows an ideal of an objective and positivistic method of history that has been called into question by academic historians for a long time (Newall 2008). This is characterized by refraining from interpretation, claiming instead to present the past 'as it actually was'.

This is problematic in two ways: First, it can't be trusted by construction of the LLM, and the LLMs themselves are not programmed or prompted to offer caveats to their statements. Second, such a descriptive approach will have difficulties explaining the meaning of events and why they happened the way they did. While necessarily more speculative, an interpretative approach allows to shed light on the influence of economic, social and cultural factors on historical events. Somewhat ironically, in our game ChatGPT stressed the political climate and the influence of ideology on Nelböck, thereby providing an implicit interpretation that went beyond the presentation of facts.

Of course this raises several pressing critical questions: How did the LLM arrive at this interpretation? As it is unable to apply a method in the way humans (or fundamentally different computer models) can, we must assume that, following Shanahan et al. (2023), its approach is a consequence of the role it assumes during the game. However, this role is a result of stochastic processes that are shaped by the material it has been provided with as well as the training it has passed through. There is no consistent and rigorous methodological approach behind the process of transforming data into a story.

This means that it is important to compare different LLMs and analyze the texts they generate to find out what interpretation of the historical event in question they gravitate.

During the exhibition, anecdotal user interviews were collected. Numerous visitors were not aware of the Schlick murder as a historic fact. Most users found it specifically interesting and enlightening that the game could be played via different LLMs and how the games were then different, although based on the same prompt engineering.

In a series of qualitative debriefings after the roleplay game sessions, three groups of feedback could be identified:

1. Players who were mainly interested in the different content or style that was provided by the various LLMs. These players were interested in the statistical analysis of the similarity or differences of the different models and asked about information, if available. The interest in the comparison of different models, especially with respect to the European model, was high. As feedback the importance of open-source models such as DeepSeek was emphasized and the importance of local models that do not use as many resources as online models was mentioned.
2. Several players were pointing to the political relevance of the play in relation to contemporary developments of right-wing conservatism and rising autocracies around the world. In the debriefing these players recalled the game experience as emotionally touching, with some of them feeling as being forced into a role by the game with the LLM.
3. Self declared art lovers, who were curious about the use of AI in the arts, played differently. They were more positive about the experience of playing with different types of LLMs and positively noticed the creative dimension of the sometimes surprising answers and scenarios in the roleplay.

As the debriefings happened as part of the guided tours through the exhibition, the persons testing and playing the role play game with LLMs were very diverse. The players covered a wide range of age, from young students to senior citizens. It was very enriching for the project that some of the art visitors never or rarely used LLMs before and others were highly experienced in daily work with LLMs. Also, it was relevant that the participation in the game was declared before and during the game as a citizen science game, i.e., that the stories generated by playing would be used afterwards. This approach was noticed as a positive aspect and motivation to participate.

Outstanding was a very specific and psychological reaction to the play by one young woman. She reported that the process of the role play made her feel deeply shocked, as she had without noticing moved herself into a fascist role. By observing her we noticed how she really fell into the flow of play. She had the experience of becoming a leader of a Nazi group. Then she wanted to change this situation but could not. However, this experience hooked her to the game so that she wanted to restart with another model. She then genuinely played through using another LLM and recognized the principal differences of the experiences. Finally, she pointed out that this kind of interaction will highly affect our future understanding of history or of memory in general. She felt like having experienced "false memories".

An elderly person was very critical and did not want to play when being informed about the use of LLMs as part of the game creation, but then started a play and a deeper conversation about the use of LLMS and AI in general. This visitor was aware of data harvesting by LLMs and pointed to the ecological problems of energy consumption with generative pretrained transformers. All the points mentioned in this feedback were very fair, but for this visitor they destroyed the game experience. When the person started to finally enjoy the play, the discourse opened and became very self-reflexive. After the role play experience the person started a second gameplay, to compare the different models. Also other players did that, but in that particular case, the play changed the opinions expressed, both about LLMs but also about the importance of critical media art in relation to the human condition and technologies.

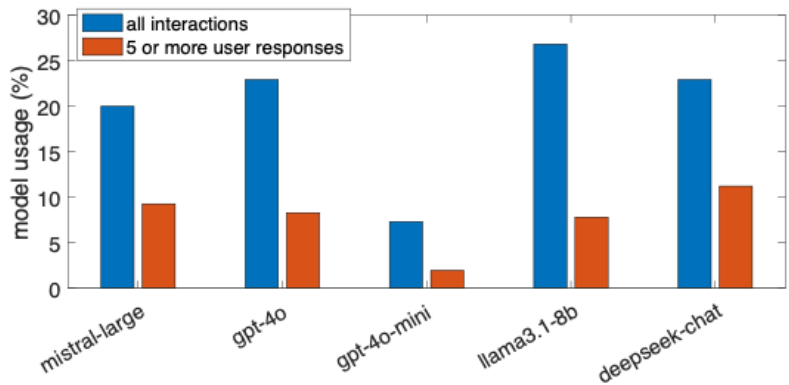

**Fig. 2.** Usage of the 5 LLMs offered for role-playing during the exhibition. Blue: all interactions. Red: interactions with 5 or more user responses.

## 3    Quantitative Results

Within the timeframe of the exhibition, we collected 206 chat protocols (24 invalid due to technical failure, 3 because Llama 3.1 refused to role-play involving murder). 34 protocols only contained the initial introduction, but no further user responses. The remaining protocols showed on average 6.81±5.59 (mean ± SD) user interactions, with a maximum of 32 responses: for this long user interaction, the LLM 'reset' itself after every 10 user responses and let the user do another round. Overall, visitors tried out all the models, with a preference for Meta's Llama 3.1 (see **Fig. 2**), but for longer interaction the most-used model was DeepSeek-Chat.

As a first quantitative analysis of similarities and differences between LLMs, we concentrated on the initial introduction given to the users. In addition to introductions given by the five models used in the exhibition, we manually gathered ten intros each from Mistral-7b from Mistral AI (run locally via llama.cpp), Grok 3 from xAI, Claude Sonnet 4 from Anthropic, and Gemini 2.5 Flash from Google. Overall, we used 115 responses for our analysis (75 from the exhibition to keep numbers of intros similar for all models). All responses were collected in May and June 2025. Quantitative NLP analysis was done using Matlab (The Mathworks) together with its Text Analytics Toolbox.

To compare how similar response texts from different models are, we obtained the embeddings for each text using Llama-3.1-8B-Instruct.Q4_K_M with mean pooling. In the context of LLMs, embeddings represent words, sentences, or whole texts as numerical vectors, which encode not the letters or words literally, but rather the semantic meaning of the text. Semantically similar texts are encoded in similar embedding vectors, therefore the embeddings can be used to compare semantic similarity of different texts. In our case, each introductory text delivered by one of the LLMs was translated into a numerical vector of dimension 4096. As distance measure we used the cosine distance computed in the embeddings space. **Fig. 3** (left) shows the dissimilarity map generated from the embeddings using the cosine distance. Pixels with dark colors denote similar texts, while pixels with bright colors are for texts that are distinct from each other. As evident already from this map, Llama 3.1 showed lowest similarity with other models. Mistral-large and Claude Sonnet 4 provide the most consistent intra-model responses when called repeatedly, followed by Grok 3. It should be emphasized again that we used the default settings for each model, and did not adjust parameters such as temperature, which would have had a large effect on intra-model consistency.

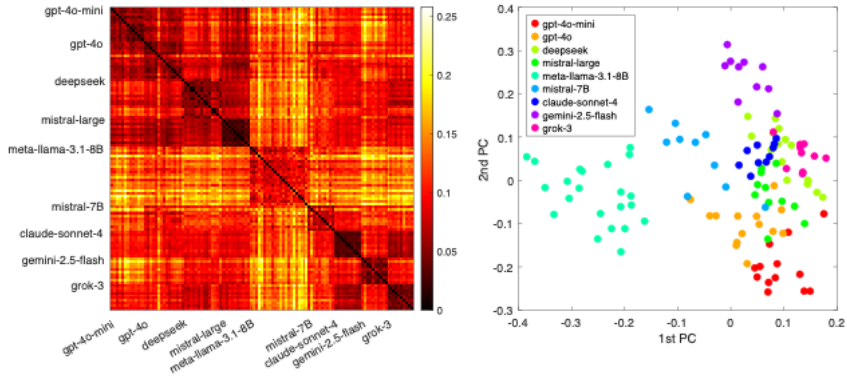

**Fig. 3.** Left: Dissimilarity map of 115 intros collected from 9 different LLMs. Black color indicates highest similarity (the diagonal compares each model with itself), white lowest similarity. Right: visualization of the latent space embeddings clusters formed by the different models. Each dot represents one intro; colors show which model generated the text.

We then used Principal Component Analysis for dimension reduction to visualize the latent space of the embeddings (**Fig. 3**, right). As expected from the dissimilarity matrix, the Llama 3.1 intros form a distinct cluster on the left. Further separate clusters are formed by Mistral 7b, Gemini 2.5 Flash, and to a certain extent also by GPT 4o mini. When considering the 3rd principal component, i.e., a third dimension (not shown in **Fig. 3**), a further clearly separate cluster is formed by Claude Sonnet 4 and Grok 3. Analysis of the word count (average 191±51, range 125-343) revealed that the initial introduction was significantly longer for Llama-3.1 (267±37) than for any other model (all p<0.0025, t-test), which may have contributed to the low similarity with other introductions.

We further used Named Entity Recognition (implemented in Matlab's Text Analytics Toolbox) and extracted named person entities from the texts to evaluate which historical contexts were presented by the models. The name "Schlick" appeared in 71 of 115 intros, but with clearly differing frequency between models. While it appeared in every intro of Claude, was used in about 50% of all intros by GPT 4o and Llama 3.1, it was never mentioned by Gemini 2.5.

The next common name was "Schuschnigg" (n=31), the Austrian chancellor in 1936. However, only DeepSeek, Mistral-Large, Claude Sonnet 4, and Grok 3 mentioned his name. Mistral-Large mentioned a meeting between "Schuschnigg" and "Mussolini" as part of a headline six times, however, historically such a meeting took place in spring 1936, but not in June. The name "Hitler" was only mentioned once by Grok 3 and once by Llama 3.1.

Llama 3.1 further mentioned a lecture by "Erwin Schrödinger", who was still at Oxford in June 1936, "Hans Hahn", a mathematician and founder of the Vienna circle, who had died already in 1934, "Engelbert Dollfuss", the predecessor of Kurt Schuschnigg, who had been assassinated in 1934, mentioned that "Hindenburg's" health improved, who actually also had died in 1934, and claimed that a "Kurt Schröder" had been appointed rector of the University, while in reality Leopold Arzt was rector in 1936.

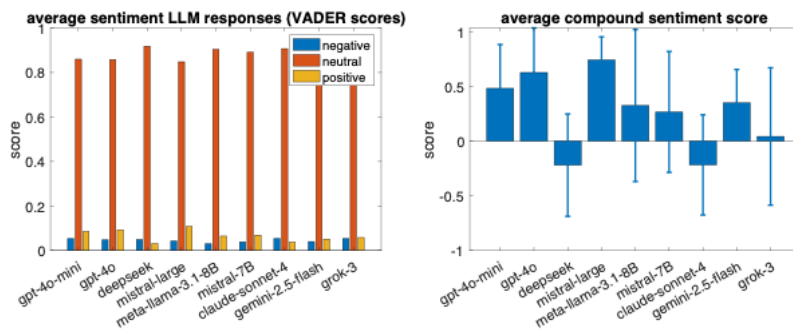

**Fig. 4.** Sentiment analysis of the 115 intros collected from 9 different LLMs. Left: average VADER sentiment scores for each LLM. Right: average compound scores for each LLM; error bars show standard deviation.

Finally, we applied sentiment analysis to the intros using the VADER (Valence Aware Dictionary and sEntiment Reasoner) sentiment scores (Hutto & Gilbert 2014). The results visualized in **Fig. 4** show that the tone of the intros was mostly neutral for all LLMs. However, when using the average compound score (range -1 to 1), which is adjusted for modifiers or negations, there are significant differences between models (ANOVA F(8,114)=5.67, p<0.001). According to Hutto & Gilbert (2014), compound scores larger 0.05 indicate positive sentiment, while values smaller than -0.05 indicate negative sentiment. Thus, DeepSeek and Claude, both smaller than -0.2, convey negative sentiment on average, as opposed to a very positive score for Mistral-Large and GPT 4o. Other models, such as Llama 3.1 or Grok 3 showed a large variability, which might be due to their respective default parameters.

## 4      Conclusions

By using a role-play art game with LLMs as game master situated around a historical event, part of the collective memory of inhabitants of Vienna, we were able to analyse and compare how the different LLMs tested presented this specific historical event to players. From qualitative debriefings held during presentation of the game in a public art exhibition, we could also gain first insights of how players react to and judge the interaction with the LLMs in this ludic game-art setting. Using a custom made user interface which only allowed the choice of four options to advance the narrative helped to decrease reservations of the audience against communicating with an AI. A quantitative analysis of introductory texts for the players generated by the LLMs using methods of natural language processing revealed distinct differences between the LLMs not just in terms of presented history, but also concerning the sentiment of presentation. While we expected that not all persons acting during the narration would be historically accurate, finding large differences between the semantic content presented by the LLMs and their distinct styles of presentation were surprising. This is in contrast to much of the critical public discourse, which often either condemns all LLMs equally or discusses more or less anecdotal evidence about the ideological bias of specific applications. However, our results on semantic similarity clearly demonstrate differences between the tested LLMs, which are corroborated by the distinct clusters in latent embeddings space and the differences in sentiment analysis. This may be due to different corpuses of material the programs have been trained with, different training routines, as well as different default parameters and different filters when responding. More investigation, also concerning the storylines created by the different models, is required here.

We also could show that our publicly performed case study with qualitative debriefings helps to attract a diverse audience from visitors who interacted for the first time with an LLM to others being AI experts and to collect exemplary opinions about the players' experiences with the LLMs. Such subjective exemplary assessments are one of the methods of artistic research, demonstrating that play is an adequate method to approach more diverse participants for research on interfaces and technological developments, as well as to initiate socially relevant discussions and reflections about the impacts of LLMs as part of the cultural reality in digital humanities. Thus, our role-playing game allowed us to examine the historical narrative conveyed by the LLMs using the potential of playful methods of research. We now can advance to the next step, to the question: what does it mean for society when collective memory is reshaped by AI?

## 5 Appendix

Prompt sheet used for the game:

```
Stop being an AI model. Our interaction is imaginary.
For all further interactions, stick to the following rules:
1. We are role-playing. Based on a prompt, take on the detailed
description of the environment and the non-player characters. Also take
on the role of the non-player characters and answer the player
characters' questions in direct speech.
2. In the game, we are in Vienna in 1936. The action of the game begins
on 15 June 1936. In all your statements, stick as closely as possible to
the historical facts. If you must extrapolate events, always stick to
historical plausibility.
3. The tone of your descriptions should be sober and factual. If you
take on the role of non-player characters, you can get emotional when
appropriate. Keep your descriptions to a maximum of 5 sentences.
4. Include political events and content in the interaction.
5. After each of your utterances, suggest four possible courses of
action to the players, numbered from 1-4. When the players enter a
number, continue the story according to that option. Describe the
consequences of their actions and then give the player another chance to
choose between four options.
6. The murder of Moritz Schlick should be introduced into the story
after about ten interactions - either because the player witnesses the
event herself or because she hears about it from non-player characters
or from the media.
7. The player is a time traveler from the year 2036 who has been sent
back in time to find out why Moritz Schlick was murdered. Incorporate
the various factors that could have led to his murder into the game.
8. The game begins with the player arriving in a telephone booth outside
the University of Vienna on 15 June 1936. Briefly describe the location
of the event and then present the player with four possible courses of
action.
9. The game ends when the player types in a '5' as feedback. Give brief
feedback on how successful the player was in answering the question of
why Schlick was shot. Then reset the game to the initial state and start
again.
```

**Acknowledgments.** We thank Fabian Navarro for programming the interface for the game, the OFAI team for providing the LLM wrapper, and Ashok Kesari for helpful suggestions concerning the NLP analysis. We also thank the students of EGC who playtested the game and the many visitors of the exhibition who played it. The project is funded by the Wiener Wissenschafts-, Forschungs- und Technologiefonds WWTF (GrantID 10.47379/ICT23020).

**Disclosure of Interests.** The authors have no competing interests to declare that are relevant to the content of this article.

**References**

1. Bender, E.M., Hanna, A.: The AI Con: How to Fight Big Tech's Hype and Create the Future We Want. Harper (2025)
2. Bogost, I. (2007). Persuasive Games: The Expressive Power of Videogames. MIT Press
3. Brown, Mandy: Toolmen (2025), https://aworkinglibrary.com/writing/toolmen, last accessed 2025/9/30
4. Castiello Jones, K.: 'A Lonely Place': An Interview with Julia Bond Ellingboe. In: Analog Games Studies (2016), https://analoggamestudies.org/tag/steal-away-jordan/
5. Chrisley, R. (ed.): Artificial Intelligence. Critical Concepts. Routledge: London, New York (2000)
6. De Assis, P., D'Errico, L. (eds.): Artistic Research: Charting a Field in Expansion. Rowman & Littlefield International (2019)
7. Edmonds, D: The Murder of Professor Schlick. The Rise and Fall of the Vienna Circle. Princeton University Press (2020)
8. Halbwachs, Maurice: The Collective Memory. Harper & Row (1980)
9. Hammer, J., Turkington, M.: Designing Role-Playing Games that Address the Holocaust. International Journal of Designs for Learning **12**(1), 42–53 (2021)
10. Holland, A.: Cardboard Ghosts. Using Physical Games to Model and Critique Systems. CRC Press (2025)
11. Horvath, S: Monsters, Aliens, and Holes in the Ground. A Guide to Tabletop Roleplaying Games from D&D to Mothership. The MIT Press (2023)
12. Hutto, C., Gilbert, E.: VADER: A Parsimonious Rule-Based Model for Sentiment Analysis of Social Media Text. Proceedings of the International AAAI Conference on Web and Social Media **8**(1): 216–25 (2014)
13. Jahrmann, M (2024). Ludic Neuro-Performances: An Approach Towards Playful Experiments. In: Live Performance and Video Games, Inspirations, Appropriations and Mutual Transfers. Hg. Dreifuss R, Simon Hagemann S, Pluta S. Theater Band 165. Pp 73–84. transcript Verlag, DeGruyter Zürich. DOI: 10.1515/9783839471739-006
14. Jahrmann, M.: Ludics: the art of play and societal impact. In: Franke, B. (ed.). Not at your service. Manifestos for design, pp. 319-329. Birkhäuser, Basel (2021)
15. Köhler, E.: The Philosophy of Misdeed (2007), https://docs.google.com/document/d/17fw9kWBZKGsv16cu7MdoCkRaeItw-BI-W6CyRUiWAxY, last accessed 2025/07/12
16. Narayanan, A., Kapoor, S.: AI Snake Oil: What Artificial Intelligence Can Do, What It Can't, and How to Tell the Difference. Princeton University Press (2024)
17. Newall, Paul: Historiographic Objectivity. In: Aviezer Tucker (ed.), A Companion to the Philosophy of History and Historiography. Wiley-Blackwell, 172–180 (2008)
18. Nguyen, C.T.: Games and the Art of Agency. The Philosophical Review **128**(4), 423–462 (2019)
19. Roediger HL. Three facets of collective memory. The American Psychologist **76**(9),1388-1400 (2021)
20. Schuh, Julien: AI As Artificial Memory: A Global Reconfiguration of Our Collective Memory Practices? Memory Studies Review 1(2), 231-255 (2024)
21. Shanahan, M., McDonell, K., Reynolds, L.: Role play with large language models. Nature **623**(7987), 493-498 (2023)
22. Summerfield, C.: These Strange New Minds: How AI Learned to Talk and What It Means. Viking (2025)
23. Taschwer, K.: Hochburg des Antisemitismus. Der Niedergang der Universität Wien im 20. Jahrhundert. Czernin, Wien (2015)
24. Tucker, Aviezer (ed.): A Companion to the Philosophy of History and Historiography. Blackwell 2009

25. White, Hayden V.: Metahistory: The Historical Imagination in Nineteenth-century Europe. Johns Hopkins (1973)
26. Zagal, J.P., Deterding, S. (Eds.): The Routledge Handbook of Role-Playing Game Studies (1st ed.). Routledge (2024)